\documentclass[twocolumn,prl,aps,showpacs,superscriptaddress,nofootinbib]{revtex4}

\usepackage{amsmath}
\usepackage{graphicx}
\usepackage{hyperref}
\usepackage{xspace}

\makeatletter
\gdef\@fpheader{}
\g@addto@macro\bfseries{\boldmath}
\makeatother

\hypersetup{
    bookmarks=true,         % show bookmarks bar?
    unicode=false,          % non-Latin characters in Acrobat’s bookmarks
    pdftoolbar=true,        % show Acrobat’s toolbar?
    pdfmenubar=true,        % show Acrobat’s menu?
    pdffitwindow=false,     % window fit to page when opened
    pdfstartview={FitH},    % fits the width of the page to the window   
    pdfnewwindow=true,      % links in new window
    colorlinks=true,        % false: boxed links; true: colored links
    linkcolor=red,          % color of internal links
    citecolor=cyan,         % color of links to bibliography
    filecolor=magenta,      % color of file links
    urlcolor=blue,         % color of external links
    linktocpage=true
}

\newcommand{\letterSec}[1]{\textbf{#1}\\}

\newcommand{\order}[1]{\mathcal{O}\!\left(#1\right)}

\newcommand{\dd}{\mathrm{d}}
\newcommand{\ee}{e}

\newcommand{\sss}[1]{{\scriptscriptstyle{#1}}}

\newcommand{\uPl}{\mathrm{Pl}}
\newcommand{\uin}{\mathrm{in}}

\newcommand{\uend}{\mathrm{end}}

\newcommand{\ucl}{\mathrm{cl}}

\newcommand{\uS}{\mathrm{S}}

\newcommand{\usssS}{\sss{\uS}}

\newcommand{\usssPl}{\sss{\uPl}}

\newcommand{\nS}{n_\usssS}

\newcommand{\uNL}{\mathrm{NL}}

\newcommand{\calP}{\mathcal{P}}

\newcommand{\Mp}{M_\usssPl}

\newcommand{\fnl}{f_\uNL}

\newcommand{\efolds}{$e$-folds~}

\newcommand{\beq}{\begin{equation}}
\newcommand{\eeq}{\end{equation}}
\newcommand{\bea}{\begin{eqnarray}}
\newcommand{\eea}{\end{eqnarray}}

\newlength{\wsingfig}
\newlength{\wdblefig}
\newlength{\wquadfig}
\newlength{\wtriplefig}
\newcommand{\Eq}[1]{Eq.~(\ref{#1})}
\newcommand{\Eqs}[1]{Eqs.~(\ref{#1})}
\newcommand{\Fig}[1]{Fig.~{\ref{#1}}}

\newcommand{\Ref}[1]{Ref.~{\cite{#1}}}

\begin{document}

\title{Critical Number of Fields in Stochastic Inflation}

\author{Vincent Vennin}
\email{vincent.vennin@port.ac.uk}
\affiliation{Institute of Cosmology \& Gravitation, University of Portsmouth, Dennis Sciama Building, Burnaby Road, Portsmouth, PO1 3FX, United Kingdom}

\author{Hooshyar Assadullahi}
\email{hooshyar.assadullahi@port.ac.uk}
\affiliation{Institute of Cosmology \& Gravitation, University of Portsmouth, Dennis Sciama Building, Burnaby Road, Portsmouth, PO1 3FX, United Kingdom}
\affiliation{School of Earth and Environmental Sciences, University of Portsmouth, Burnaby Building, Burnaby Road, Portsmouth, PO1 3QL, United Kingdom}

\author{Hassan Firouzjahi}
\email{firouz@ipm.ir}
\affiliation{School of Astronomy, Institute for Research in Fundamental Sciences (IPM), Tehran, Iran, P.O. Box 19395-5531}

\author{Mahdiyar Noorbala}
\email{mnoorbala@ut.ac.ir}
\affiliation{Department of Physics, University of Tehran, Iran, P.O. Box 14395-547}
\affiliation{School of Astronomy, Institute for Research in Fundamental Sciences (IPM), Tehran, Iran, P.O. Box 19395-5531}
 
\author{David Wands}
\email{david.wands@port.ac.uk}
\affiliation{Institute of Cosmology \& Gravitation, University of Portsmouth, Dennis Sciama Building, Burnaby Road, Portsmouth, PO1 3FX, United Kingdom}

\date{\today}

\begin{abstract}
Stochastic effects in generic scenarios of inflation with multiple fields are investigated. First passage time techniques are employed to calculate the statistical moments of the number of inflationary $e$-folds, which give rise to all correlation functions of primordial curvature perturbations through the stochastic $\delta N$ formalism. The number of fields is a critical parameter. The probability of exploring arbitrarily large-field regions of the potential becomes non-vanishing when more than two fields are driving inflation. The mean number of $e$-folds can be infinite, depending on the number of fields; for plateau potentials, this occurs even with one field. In such cases, correlation functions of curvature perturbations are infinite. They can, however, be regularised if a reflecting (or absorbing) wall is added at large energy or field value. The results are found to be independent of the exact location of the wall and this procedure is, therefore, well-defined for a wide range of cutoffs, above or below the Planck scale. Finally, we show that, contrary to single-field setups, multi-field models can yield large stochastic corrections even at sub-Planckian energy, opening interesting prospects for probing quantum effects on cosmological fluctuations.
\end{abstract}

\maketitle

\letterSec{Introduction}
In the inflationary paradigm~\cite{Starobinsky:1980te, Sato:1980yn, Guth:1980zm, Linde:1981mu, Albrecht:1982wi, Linde:1983gd}, cosmological inhomogeneities result from the parametric amplification of the vacuum quantum fluctuations of the gravitational and matter fields during an early accelerated expansion~\cite{Starobinsky:1979ty, Mukhanov:1981xt, Hawking:1982cz,  Starobinsky:1982ee, Guth:1982ec, Bardeen:1983qw}. The transition from quantum fluctuations to classical but stochastic density perturbations~\cite{Polarski:1995jg,Lesgourgues:1996jc,Kiefer:2008ku,Martin:2012pea,Burgess:2014eoa,Martin:2015qta} plays an important role in this scenario. In particular, it implies that the open quantum system comprising the super-Hubble degrees of freedom can be described with a classical stochastic theory, the stochastic inflation formalism~\cite{Starobinsky:1982ee, Starobinsky:1986fx, Nambu:1987ef, Nambu:1988je, Kandrup:1988sc, Nakao:1988yi, Nambu:1989uf, Mollerach:1990zf, Linde:1993xx, Starobinsky:1994bd}. This consists of an effective description of the long-wavelength parts of the quantum fields, which are ``coarse grained'' at a fixed physical scale, larger than the Hubble radius during the whole inflationary era. In this framework, the short-wavelength quantum fluctuations act as a classical noise on the dynamics of the super-Hubble scales, and at leading order in slow roll, the coarse-grained fields $\phi_i$ follow Langevin equations
\begin{align}
\label{eq:Langevin}
\frac{\dd\phi_i}{\dd N}=-\frac{1}{3H^2}\frac{\partial V}{\partial\phi_i}+\frac{H}{2\pi}\xi_i\, ,
\end{align}
labeled by the number of \efolds $N$, where $V$ is the inflationary potential, $H\simeq\sqrt{V/(3\Mp^2)}$ is the Hubble parameter, and $\xi_i$ are independent normalised white Gaussian noises satisfying $\langle\xi_i(N_1)\xi_j(N_2)\rangle = \delta_{i,j}\delta(N_1-N_2)$.

Combined with the $\delta N$ formalism, this allows one to study how quantum effects modify inflationary observable predictions. In \Ref{Vennin:2015hra}, such a ``stochastic $\delta N$ formalism'' was used to show that, even if stochastic effects can shift the location of the observational window along the inflationary potential, in single-field inflation, stochastic corrections within the observational window are always small at sub-Planckian energy. In this Letter, we build on the results of \Ref{Assadullahi:2016gkk} to investigate the generic situation where $D$ scalar fields are driving inflation.\\

\letterSec{The stochastic $\delta N$ formalism}
In the $\delta N$ formalism~\cite{Starobinsky:1982ee, Starobinsky:1986fxa, Salopek:1990jq, Sasaki:1995aw, Sasaki:1998ug, Wands:2000dp, Lyth:2004gb, Lyth:2005fi}, scalar curvature perturbations $\zeta$ are identified with fluctuations in the number of \efolds $N$ realised between an initial flat slice of space-time and a final slice of uniform energy density, among a family of homogeneous universes. In the stochastic picture, this number of \efolds is a random variable that we denote $\mathcal{N}$, and its statistical moments thus directly give rise to the correlation functions of cosmological perturbations. 

For example, the power spectrum of curvature perturbations (two-point correlation function) is related to the second moment of $\mathcal{N}$ and can be expressed as
\begin{align}
\calP_\zeta=\dfrac{\dd\left(\left\langle \mathcal{N}^2 \right\rangle - \left\langle \mathcal{N} \right\rangle^2\right)}{\dd\left\langle \mathcal{N} \right\rangle}\, ,
\label{eq:Pzeta:Nmean}
\end{align}
where $\mathcal{N}$ is the number of \efolds realised between the time when the scale at which $\calP_\zeta$ is calculated exits the Hubble radius during inflation and the end of inflation (where the power spectrum is calculated). In the same manner, the local $\fnl$ parameter (three-point correlation function), measuring the ratio between the bispectrum and the power spectrum squared, is given by
\begin{align}
\fnl=\frac{5}{72}
\frac{\dd^2\left\langle\left(\mathcal{N}-\left\langle\mathcal{N}\right\rangle\right)^3 \right\rangle}
{\dd\left\langle\mathcal{N}\right\rangle^2}
\calP_\zeta^{-2}\, ,
\label{eq:fNL:Nmean}
\end{align}
and higher moments can be expressed with analogous expressions.
This is the so-called ``stochastic $\delta N$ formalism''~\cite{Enqvist:2008kt, Fujita:2013cna, Fujita:2014tja, Vennin:2015hra, Kawasaki:2015ppx, Assadullahi:2016gkk}. 
%In fact, this approach may also be called ``stochastic-$N$ formalism'' since it does not rely on an expansion in $\delta N$ and in the metric perturbation $\zeta$ (for instance, these two quantities are not small in the so-called regime of ``eternal inflation''). 
It reduces the problem to calculating the statistical moments of the number of \efolds realised under \Eq{eq:Langevin}.

This can be performed using first passage time analysis~\cite{Bachelier:1900, Gihman:1972}. Starting from an initial point $\phi_i^\uin$ in field space, the set of functions $f_n(\phi_i)\equiv\langle\mathcal{N}^n\rangle(\phi_i^\uin=\phi_i)$ are shown to satisfy the set of (deterministic) partial differential equations~\cite{Vennin:2015hra, Assadullahi:2016gkk}
\begin{align}
\sum_i\left(v\frac{\partial^2}{\partial\phi_i^2}-
\frac{v_{\phi_i}}{v}\frac{\partial}{\partial\phi_i}\right)f_n= -n\frac{f_{n-1}}{\Mp^2}\, .
\label{eq:PDE}
\end{align}
These equations are valid for $n\geq 1$, where we have defined $f_0=1$ and $v=V/(24\pi^2\Mp^4)$. They must be solved according to some boundary conditions. One of them is provided by the requirement that the $f_n$ functions vanish on the field-space hypersurface $\partial\Omega_-$ where inflation ends (and where the correlation functions are calculated). In case the inflationary field-space domain $\Omega$ is compact, as in hilltop models, this is sufficient to define a Cauchy problem. In other situations however, where inflation can proceed at arbitrarily large-field values, this single absorbing condition on $\partial\Omega_-$ is not enough and a second reflecting (or absorbing) boundary condition must be placed at large-field values on $\partial\Omega_+$. Whether or not this boundary condition has an observational effect when taken to sufficiently large values is one of the main questions addressed in this work. 

The procedure one has to follow is therefore the following: solve \Eqs{eq:PDE} with the boundary conditions just discussed, and use \Eqs{eq:Pzeta:Nmean} and~(\ref{eq:fNL:Nmean}) to derive $\calP_\zeta=\dd f_2/\dd f_1-2f_1$, $\fnl=5/72\calP_\zeta^{-2}\, \dd^2(f_3-f_1^3-3f_1f_2+3f_1^3)/\dd f_1^2$, etc.\\

\letterSec{The single-field case}
If only one field is present ($D=1$), \Eq{eq:PDE} is an ordinary differential equation that can be solved exactly~\cite{Vennin:2015hra}
\begin{align}
\label{eq:fn:sol:onefield}
f_n(\phi)=n \int_{\phi_-}^\phi\frac{\dd x}{\Mp}\int^{\phi_+}_x\frac{\dd y}{\Mp}\ee^{\frac{1}{v(y)}-\frac{1}{v(x)}}\frac{f_{n-1}(y)}{v(y)}\, .
\end{align}
This gives rise to all correlation functions explicitly. The standard results can be recovered as the sub-Planckian limit $v\ll 1$ of the above formula using saddle-point approximations. For the power spectrum, for instance, one obtains~\cite{Vennin:2015hra}
\begin{align}
\calP_\zeta = \frac{2}{\Mp^2}\frac{v^3}{{v^\prime}^2}\left[1+v\left(5-4\frac{v v^{\prime\prime}}{{v^\prime}^2}\right)+\mathcal{O}\left(v^2\right)\right]\, ,
\label{Pzeta:SF:classLim}
\end{align}
where the first term corresponds to the usual classical result $\calP_{\zeta,\ucl}=2v^3/(\Mp^2{v^\prime}^2)$~\cite{Mukhanov:1985rz, Mukhanov:1988jd}, and where primes mean derivatives with respect to the inflaton field. In the same manner, the three-point correlation function can be calculated and in the sub-Planckian limit, one has
% expanded in the arXiv version
\begin{align}
\fnl&=\frac{5}{24}\Mp^2\left[6\frac{{v^\prime}^2}{v^2}-4\frac{v^{\prime\prime}}{v}
\nonumber \right. \\ & \left.
+v\left(25\frac{{v^\prime}^2}{v^2}-34 \frac{v^{\prime\prime}}{v}-10\frac{v^{\prime\prime\prime}}{v^\prime}+24\frac{{v^{\prime\prime}}^2}{{v^\prime}^2}\right)+\mathcal{O}\left(v^2\right)\right]\, ,
\label{fnl:SF:classLim}
\end{align}
%$\fnl=5\Mp^2/24[6{v^\prime}^2/v^2-4v^{\prime\prime}/v+v(25{v^\prime}^2/v^2-32 v^{\prime\prime}/v-2v^{\prime\prime\prime}/v^\prime+22{v^{\prime\prime}}^2/{v^\prime}^2)+\mathcal{O}(v^2)]$, 
where the first line again matches the standard result~\cite{Maldacena:2002vr, Kenton:2015lxa}. In contrast, note that this standard result cannot be obtained within the usual classical $\delta N$ formalism because of the intrinsic non-Gaussianity of the field at Hubble exit~\cite{Maldacena:2002vr, Allen:2005ye}.

These expressions make clear that stochastic corrections to single-field observables scale with $v$ (since they come from the self- and gravitational interaction of the inflaton field), which is always small in the observational window. For example, the leading order relative correction in \Eq{Pzeta:SF:classLim} can be written as $5v-4v^2v^{\prime\prime}/{v^\prime}^2 = \calP_{\zeta,\ucl}(1-\nS-r/16)<10^{-10}$, where $\nS$ and $r$ are the (classical) scalar spectral index and tensor-to-scalar ratio.\\

\letterSec{Infinite Inflation}
If more than one scalar field is present, \Eq{eq:PDE} is a full partial differential equation which does not have generic solutions. In order to proceed analytically, we therefore illustrate our study with the subclass of potentials $v(r)$ that depend on the radial coordinate
\begin{align}
r=\sqrt{\sum_{i=1}^D\phi_i^2}
\end{align}
only, and for which, assuming radially symmetric boundary conditions at $r_\pm$, \Eq{eq:PDE} has analytical solutions~\cite{Assadullahi:2016gkk}
\begin{align}
f_n(r)=
n\displaystyle\int_{r_-}^r\frac{\dd x}{\Mp}
\displaystyle\int_{x}^{r_+}\frac{\dd y}{\Mp}
\ee^{\frac{1}{v(y)}-\frac{1}{v(x)}}\frac{f_{n-1}(y)}{{v(y)}}
\left(\frac{y}{x}\right)^{D-1}.
\label{eq:v(r):sol}
\end{align}
When $D=1$, $r=\phi$ and one can check that the single-field solution~(\ref{eq:fn:sol:onefield}) is recovered.
This expression allows us to study the behaviour of the moments of the number of \efolds when the location of the extra boundary condition $r_+$ is removed to infinity. If the potential $v(r)$ has an asymptotic monomial profile $v(r)\propto r^p$ at large $r$, one can check that the above integrals diverge when $r_+\rightarrow \infty$ as soon as $D\geq p$. We call this phenomenon ``infinite inflation''. Note that it implies, but is not equivalent to, eternal inflation~\cite{Steinhardt:1982kg, Vilenkin:1983xq, Guth:1985ya, Linde:1986fc, Guth:2007ng} (for example, eternal inflation can be realised with a single-field quadratic potential but not infinite inflation). If the potential is of the plateau type for instance ($p=0$), one always has infinite inflation. Otherwise infinite inflation occurs when a sufficiently large number of fields are present. 

In this context, the number of fields plays the role of an order parameter, as in phase transitions or recurrence problems~\cite{Polya:1921}. Infinite inflation takes place independently of the initial field values, so that these infinities cannot be removed by simply going to sufficiently low energy. This raises two questions. First, if infinite numbers of \efolds are realised, is the system exploring very large-field regions of the potential that are classically inaccessible and to which observations should be sensitive? Second, can these infinities be regularised away from observable quantities? \\

\letterSec{Large-field exploration}
The probability $p_+(\phi_i)$ that, starting from $\phi_i^\uin=\phi_i$, the system reaches $\partial\Omega_+$ before $\partial\Omega_-$ can be shown~\cite{Assadullahi:2016gkk} to satisfy the partial differential equation
\begin{align}
\sum_i\left(v\frac{\partial^2}{\partial\phi_i^2}-
\frac{v_{\phi_i}}{v}\frac{\partial}{\partial\phi_i}\right)p_+=0\, ,
\label{eq:PDE:p_+}
\end{align}
with boundary conditions $p_+=0$ on $\partial\Omega_-$ and $p_+=1$ on $\partial\Omega_+$. When $\partial\Omega_+$ is removed to infinity, we call $p_+$ the ``large-field exploration probability''. For the $v(r)$ potentials introduced above, \Eq{eq:PDE:p_+} can be solved and one obtains
\begin{align}
\label{eq:v(r):p+}
p_+\left(r\right)=\displaystyle\dfrac{\displaystyle\int_{r_-}^r {x}^{1-D}\ee^{-\frac{1}{v({x})}}\dd {x}}{\displaystyle\int_{r_-}^{r_+} {x}^{1-D}\ee^{-\frac{1}{v({x})}}\dd {x}}\, .
\end{align}
When $r_+\rightarrow\infty$, $p_+>0$ if the function $r^{1-D}$ is integrable (assuming that $v$ is positive at infinity). Contrary to the case of infinite inflation, this is independent of the shape of the potential at large-field value, and $p_+>0$ as soon as more than $2$ fields are present. This further illustrates why the number of fields is a critical parameter in stochastic inflation.

When $D>2$, $p_+>0$ but its actual value is in fact very small if the dynamics is started at sub-Planckian energy. Indeed, substantial large-field exploration probabilities are obtained if the integrand of \Eq{eq:v(r):p+} is maximal in the numerator integration domain. If the potential has a monomial asymptotic profile $v(r)\propto r^p$ with $p>0$, this is the case if initial conditions are such that $v_\uin>p/(D-1)$. In plateau potentials where $v$ approaches a constant value $v_\infty$ at infinity, this happens if $v_\infty>\mathcal{O}(0.1)/(D-2)$. However, if one normalises the overall mass scale of the potential to fit the measured amplitude~\cite{Ade:2015xua} of the scalar power spectrum and starts the evolution $50$ (classical) \efolds before the end of inflation, one has $v_\uin \sim  10^{-11}p$ for monomial potentials and $v_\infty \sim 10^{-12 }$ for plateau potentials, so that $10^{11}$ fields would be required to obtain appreciable values for $p_+$. Therefore, even though multiple fields yield non-vanishing large-field exploration probabilities, in practice, sub-Planckian energies prevent these probabilities from being non-negligible, and ``protect'' the dynamics from the large-field regimes of the theory.\\

\letterSec{Regularisation}
\begin{figure*}
\begin{center}
\includegraphics[width=0.49\textwidth]{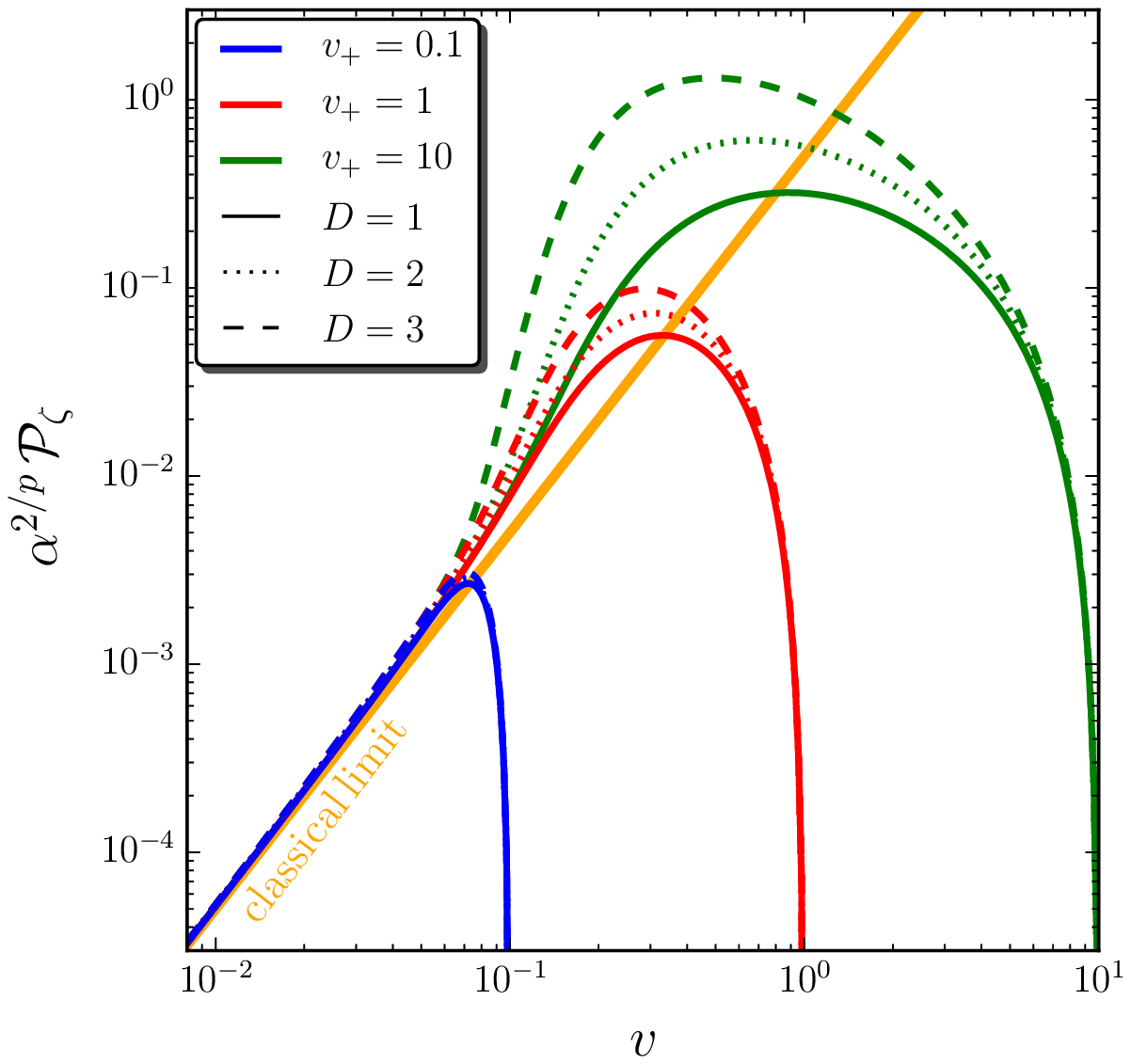}
\includegraphics[width=0.49\textwidth]{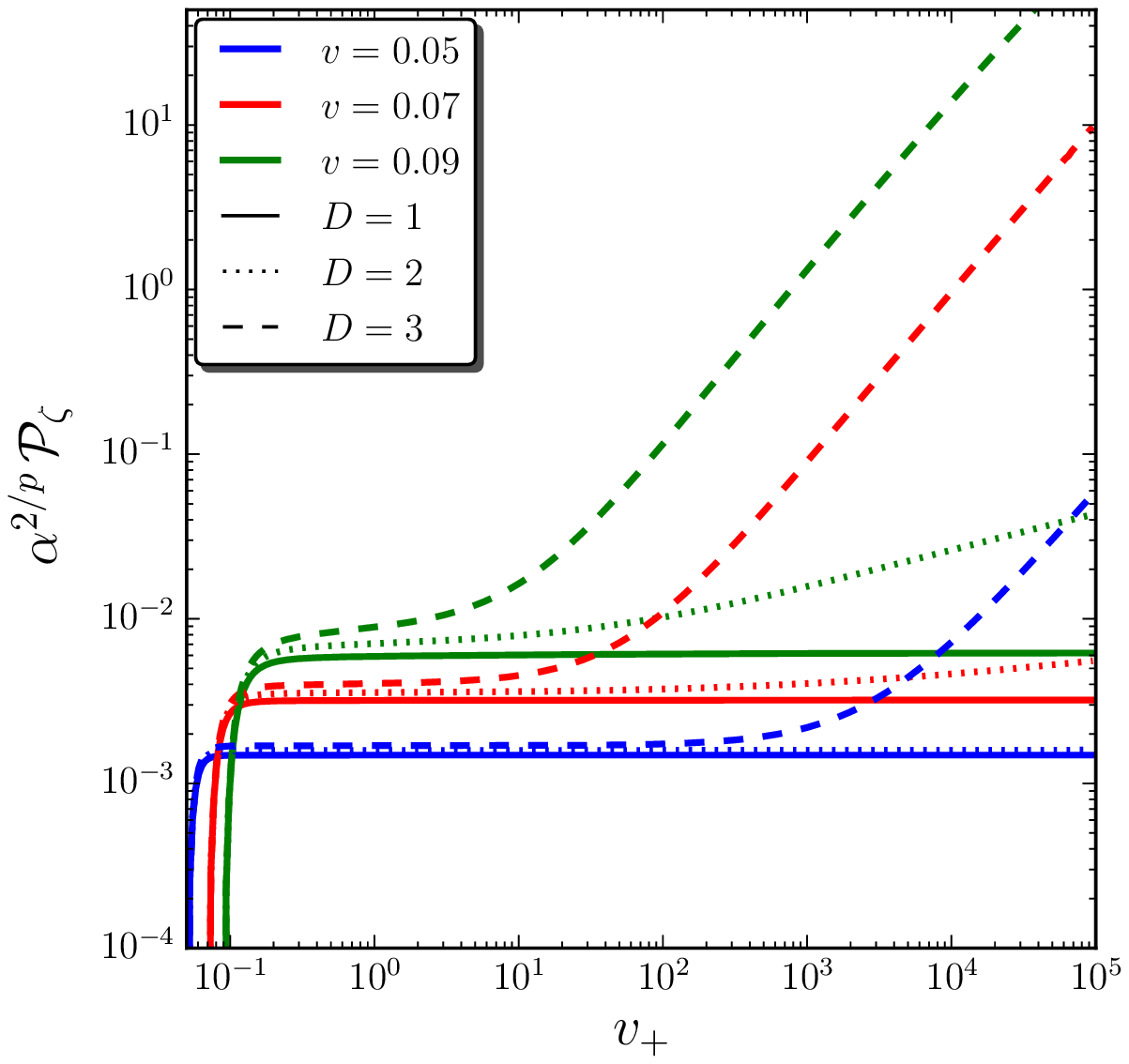}
\caption{Scalar power spectrum amplitude $\calP_\zeta$ in $v = \alpha \sum_{i=1}^D\phi_i^2$ potentials, as a function of the potential energy $v$ at which the scales for which  $\calP_\zeta$ is calculated exit the Hubble radius during inflation (left panel), and as a function of the upper reflecting wall location $v_+$ in the right panel, for a few values of the number of fields $D$.}
\label{fig:Pzeta}
\end{center}
\end{figure*}
Let us see if a similar mechanism exists for the observable quantities of the problem; namely, the correlation functions of scalar adiabatic perturbations. Combining \Eqs{eq:Pzeta:Nmean} and~(\ref{eq:v(r):sol}), the scalar power spectrum $\calP_\zeta$ in $v(r) = \alpha r^2$ potentials is plotted in \Fig{fig:Pzeta} for illustrative purposes. In the left panel, $\calP_\zeta$ is displayed as a function of $v$ (where the scale at which the power spectrum is calculated exits the Hubble radius), for a few values of $D$ and $v_+$ (where a reflecting wall is located). In the sub-Planckian limit where $v \ll 1$, all curves approach the classical formula $\calP_{\zeta,\ucl}=v^2/(2\alpha\Mp^2)$, which is independent of the number of fields $D$. When $v$ is of order $0.1$ or greater, the full result deviates from the classical prediction, in a way that depends on $v_+$ and $D$. Therefore, stochastic effects introduce dependences on these parameters that do not exist in the classical picture otherwise. Let us discuss the role played by both quantities.

In the right panel of \Fig{fig:Pzeta}, $\calP_\zeta$ is displayed as a function of $v_+$ for a few values of $v$ and $D$. When $D=1$ (solid lines), $\calP_\zeta$ converges to a finite value when $v_+\rightarrow\infty$, which is reached soon after $v_+\gg v$ and therefore provides a well-defined prediction when the reflecting wall is removed to infinite energy. However, when $D\geq 2$ (dashed and dotted lines), $\calP_\zeta$ diverges when $v_+\rightarrow\infty$, as a consequence of the phenomenon of infinite inflation discussed below \Eq{eq:v(r):sol}. In this case, a reflecting (or absorbing) wall at large-field value is compulsory to make the power spectrum (as well as higher correlators) finite. For plateau potentials, let us recall that this happens regardless of the number of fields. 

In such cases, how much does the result depend on the precise location of this large-field wall? In the right panel of \Fig{fig:Pzeta}, one can see that when $v_+$ increases, the power spectrum amplitude reaches a plateau the width of which decreases with $v$, before diverging. More precisely, one can show that the contribution from the upper bound $r_+$ of the second integral in \Eq{eq:v(r):sol} is subdominant when
\begin{align}
v\ll v_+\ll  
\ee^{\frac{\mathcal{O}(1)}{v}}
\end{align}
for $v(r)\propto r^p$ potentials, where the $\order{1}$ constant depends on $p$ and $D$. For example, if one takes $D=2$ and $p=3$, $v_\uin\sim 10^{-10}$ leads to $v_+\ll 10^{3,474,355,825}$. This is an extremely large, ``ultra super-Planckian'' value below which quantum gravity effects are expected to come into play anyway. For plateau potentials, one finds 
\begin{align}
r\ll r_+\ll \ee^{\order{1}\left(\frac{1}{v}-\frac{1}{v_\infty}\right)}\, ,
\end{align}
where the $\order{1}$ constant depends on the exact shape of the plateau and on the number of fields. In the Starobinsky model~\cite{Starobinsky:1980te} with a single field for instance, 
%$v=v_\infty[1-\exp(-\sqrt{2/3}r/\Mp)]^2$, $50$ \efolds before the end of inflation one has $r\simeq 5.24\Mp$ and $v\sim 0.986 v_\infty$, so that $v_\infty\sim 10^{-12}$ with one single field,
one obtains $r_+ \ll 10^{6,166,453,090} \Mp$. This value is again huge and one typically expects~\cite{Broy:2014sia, Coone:2015fha} monomial corrections to spoil the plateau potential before then, which would bring us back to the previous monomial case. As a consequence, if inflation proceeds at sub-Planckian energy, predictions are independent of the location of the large-field wall, provided it is placed below the ultra super-Planckian values just quoted. 

Stochastic effects and infinite inflation therefore require modifying the super-Planckian limit of inflationary models to make them consistent, but this modification does not impact their predictions, up to corrections typically of order $\ee^{-\order{1}/v}$. If we neglect these, performing a saddle-point approximation of \Eq{eq:v(r):sol} in the $v\ll 1$ limit, the scalar power spectrum for $v(r)$ potentials is given by
\begin{align}
\calP_\zeta = \frac{2}{\Mp^2}\frac{v^3}{{v^\prime}^2}\left[1+v\left(5-4\frac{v v^{\prime\prime}}{{v^\prime}^2}+2\frac{D-1}{r}\frac{v}{v^\prime}\right)+\mathcal{O}(v^2)\right]\, ,
\label{Pzeta:v(r):classLim}
\end{align}
while the non-Gaussianity parameter $\fnl$ reads 
%%%%%%% expand in the arXived version  %%%%%%%%%%
\begin{align}
&\fnl=\frac{5}{24}\Mp^2\left\lbrace 6\frac{{v^\prime}^2}{v^2}-4\frac{v^{\prime\prime}}{v}
+v\left[25\frac{{v^\prime}^2}{v^2}-34 \frac{v^{\prime\prime}}{v}
\nonumber \right.\right. \\ & \left.\left.
-10\frac{v^{\prime\prime\prime}}{v^\prime}+24\frac{{v^{\prime\prime}}^2}{{v^\prime}^2}
+2\frac{D-1}{r^2}\left(r\frac{v^\prime}{v}-2\right)\right]
+\mathcal{O}\left(v^2\right)\right\rbrace\, .
\label{fnl:v(r):classLim}
\end{align}
When $D=1$, one recovers \Eqs{Pzeta:SF:classLim} and~(\ref{fnl:SF:classLim}).
%%%%%%%%%%%%%%%%%%%%%%%%%%%%%%%%%%%%%%%%%%%%%%%%%
%%%%%%% contracted in the PRL version  %%%%%%%%%%
%$\fnl=5\Mp^2/24\lbrace 6{v^\prime}^2/v^2-4v^{\prime\prime}/v+v[25{v^\prime}^2/v^2-34 v^{\prime\prime}/v-10v^{\prime\prime\prime}/v^\prime+24{v^{\prime\prime}}^2/{v^\prime}^2+2(D-1)/r^2(rv^\prime/v-2)]+\mathcal{O}(v^2)\rbrace$. 
%When $D=1$, one recovers \Eq{Pzeta:SF:classLim} and the expression given below for $\fnl$. 
%%%%%%%%%%%%%%%%%%%%%%%%%%%%%%%%%%%%%%%%%%%%%%%%%

In these expressions, the $D$-dependent terms are typically suppressed by $v$, as the other single-field stochastic corrections. Does it mean that, in multiple-field setups as well, stochastic corrections to correlation functions are always small in the observational window?\\

\letterSec{Large stochastic effects at sub-Planckian energy}
\begin{figure*}
\begin{center}
\includegraphics[width=0.49\textwidth]{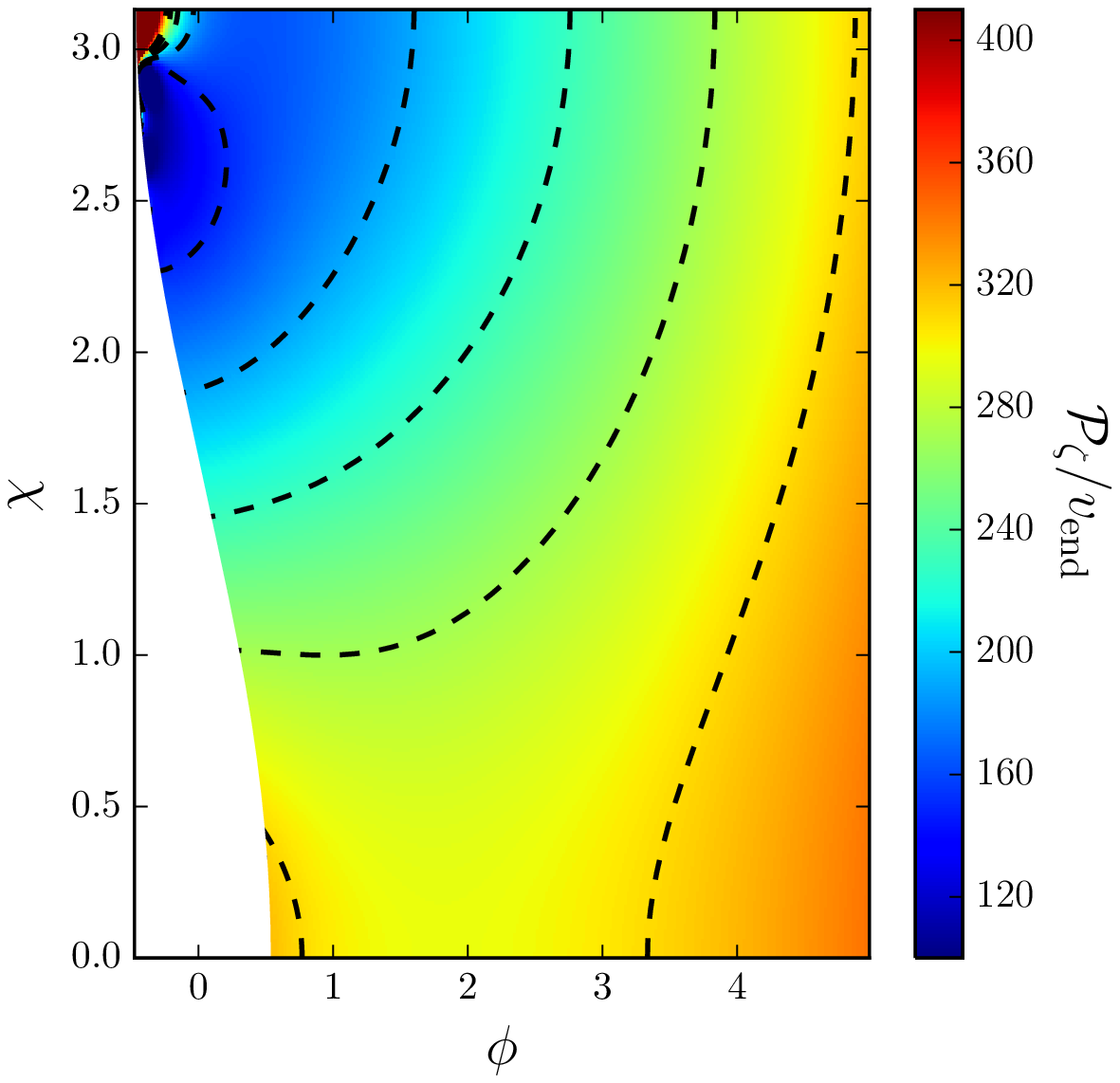}
\includegraphics[width=0.49\textwidth]{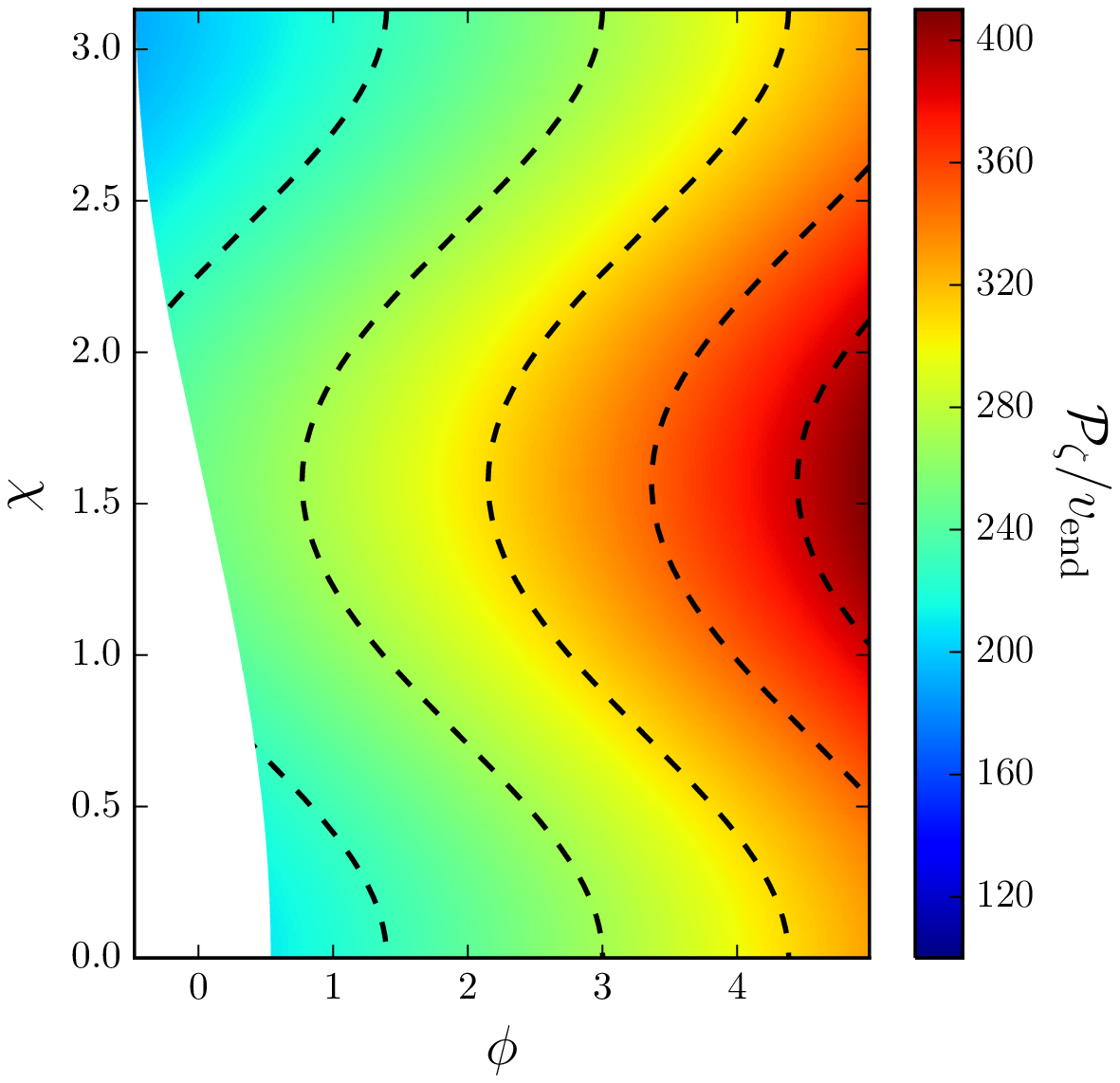}
\caption{Scalar power spectrum amplitude $\calP_\zeta$ (left panel: full stochastic results, right panel: classical formula) for the single-field potential $v=v_\uend \ee^{\alpha\phi}$ when the end of inflation is modulated by an extra field $\chi$ through $\phi_-(\chi)=\mu\cos(\chi/\chi_0)$, for $\alpha=0.1$, $\chi_0=\Mp$, $\mu=0.5\Mp$ and $v_\uend=0.05$ (this last value does not lead to the right scalar power spectrum amplitude~\cite{Ade:2015xua}, but it is used for computational convenience).
The black dashed lines are various level lines of $\calP_\zeta$, and the white regions correspond to $\phi<\phi_-(\chi)$ which is outside the inflationary domain. 
%One may also note that only the region $0\leq \chi\leq \pi\chi_0$ is displayed since the result is symmetric about $\chi=\pi\chi_0$ and $2\pi\chi_0$-periodic.
}
\label{fig:inhomoegenous}
\end{center}
\end{figure*}
The class of $v(r)$ potentials discussed so far allowed us to study field space volume effects related to the number of fields. However, these potentials are special since they are globally flat in directions orthogonal to the classical trajectory. But including more than one scalar field also opens the possibility of letting stochastic effects explore features of the potential away from the classical path. This can lead to large stochastic effects even at sub-Planckian energy.

To illustrate this property, let us consider the simple case~\cite{Dvali:2003ar} where inflation is driven by a single field $\phi$ but the surface defining the end of inflation is modulated by a second field $\chi$ through $\phi_-(\chi)$. In \Ref{Assadullahi:2016gkk}, the case where $v\propto\ee^{\alpha\phi/\Mp}$ and $\phi_-(\chi)=\mu\cos(\chi/\chi_0)$ is investigated (see Fig.~5 of this reference). It is shown that when the initial value of $\phi$ increases, $\langle\mathcal{N}\rangle$ becomes independent of the initial value of $\chi$, contrary to the classical limit where one simply has $N_\ucl=[\phi-\phi_-(\chi)]/(\Mp\alpha)$. This is because the diffusion term acting on $\chi$ randomises its \textit{vev} and tend to erase memory of its initial value. In the stochastic regime, the exit point is spread over the entire end-surface.

In \Fig{fig:inhomoegenous}, the corresponding scalar power spectrum is displayed. One can see that the amplitude is generically smaller than the classical prediction (since the additional contribution from the inhomogeneous end of inflation tends to be smeared out), but that the entire shape of the power spectrum is also substantially modified. In practice, the size of the effect depends on how the scale over which modulation takes place (denoted $\chi_0$ in the present model) compares to the dispersion acquired by the modulating field at the end of inflation (of the order of $\sqrt{2v N}\Mp$). If one sets $v\simeq 10^{-10}$ and $N\simeq 50$, one finds that the effect is large if $\chi_0 \lesssim 10^{-4}\Mp$.

It is therefore interesting to notice that, contrary to the purely single-field case~\cite{Vennin:2015hra}, large stochastic corrections can be obtained in the present situation even if $v\ll 1$, depending on the scales of the features in the end-surface. As another example, let us mention hybrid inflation, where stochastic effects play a crucial role in triggering the tachyonic instability~\cite{Martin:2011ib, Levasseur:2013tja, Kawasaki:2015ppx}.\\

\letterSec{Conclusion}
The stochastic $\delta N$ formalism is a powerful tool to calculate quantum backreaction effects on cosmological observables during inflation. When applied to multiple field scenarios, it reveals that the field space dimension plays a critical role. In particular, introducing several scalar fields generically leads to infinite moments of the number of $e$-folds, which translate into infinite correlation functions of curvature perturbations. In this Letter, we have shown how these infinities can be regularised by introducing a reflecting (or absorbing) wall at high energy or field value, and why the results are independent of the exact location of this wall, up to tiny $\ee^{-1/v}$ corrections. Another fundamental difference between single- and multiple-field inflation is that contrary to single-field setups, large stochastic effects can be found in the observational window for some multiple field scenarios even at sub-Planckian energy. This opens interesting prospects for probing quantum effects on inflationary dynamics.\\

\begin{acknowledgments}
V.V., H.A., and D.W. acknowledge financial support from STFC Grants No. ST/K00090X/1 and No. ST/N000668/1. M.N.\ acknowledges financial support from the research council of the University of Tehran.
\end{acknowledgments}

\bibliography{MultiStocha}

\end{document}